\documentclass[osajnl,twocolumn,showpacs,groupaddress,10pt]{revtex4-1} 
\usepackage{amsmath,amssymb,graphicx}
\usepackage{epstopdf}
\usepackage[utf8]{inputenc}
\usepackage{float}

\begin{document}

\title{Negative Index Materials:\\The Key to "White" Multilayer Fabry-Perot}

\author{Michel Lequime}\email{Corresponding author: michel.lequime@fresnel.fr}
\author{Boris Gralak}
\author{Sébastien Guenneau}
\author{Myriam Zerrad}
\author{Claude Amra}
\affiliation{Aix Marseille Université, CNRS, Centrale Marseille, Institut Fresnel, UMR 7249, 13013 Marseille, France}

\begin{abstract}
The use of negative index materials is highly efficient for tayloring the spectral dispersion properties of a quarter-wavelength Bragg mirror and for obtaining a resonant behavior of a multilayer Fabry-Perot cavity over a very large spectral range. An optimization method is proposed and validated on some first promising devices. 
\end{abstract}

\maketitle

\section{Introduction}
\label{sec:Introduction}
The proposal of materials with simultaneous negative electric permittivity and magnetic permeability by Veselago in 1967 \cite{Veselago_1968} has opened the door toward the design of novel and remarkable optical devices based on the use of metamaterials or photonic crystals, such as the perfect flat lens \cite{Pendry_2000} or the invisibility cloak \cite{Schurig_2006}.

Recently, we have shown how these negative electromagnetic properties can be revisited through the admittance formalism \cite{Lequime_2013}, which is widely used in the thin-film community \cite{Macleod_2010} and defined the computational rules for the effective indices and phase delays associated with wave propagation through negative-index layers \cite{Lequime_2013_2}. We have demonstrated that we can simulate the optical properties of negative index material (NIM) layer by replacing it with a positive index material (PIM) with the same effective index ($\tilde{n}_{\text{PIM}}=\tilde{n}_{\text{NIM}}$), provided that we use for this PIM layer a \textbf{virtual} thickness \textbf{opposite} to that of the NIM layer ($t_{\text{PIM}}=-t_{\text{NIM}}$), which is reminiscent of optical space folding in complementary media \cite{Pendry_2003}.

This computational rule is easily implementable in standard thin-film software and has allowed us to analyze the spectral properties of some standard multilayer stacks, such as the antireflection coating, the quarter-wavelength Bragg mirror and the Fabry-Perot bandpass filter, in which one or more layers of these stacks involve negative index materials \cite{Lequime_2013_2}.

Among the presented results, the most spectacular concerns the large increase in the spectral bandwidth of a quarter-wavelength Bragg mirror induced by the use of a negatively refracting material (either the high-index layers or the low-index layers) and the ability to tailor the phase properties of such multilayer structures by adjusting the number and the features of NIM layers within the stack.

The objective of this work is to define an optimization method for such improvements and to identify the design of a \textit{White} Fabry-Perot, i.e. a multilayer cavity that spontaneously exhibits \textbf{resonant behavior over a very large spectral range}.

\section{Quarter-wavelength Bragg mirror}

Let us consider a Bragg mirror that contains $p$ alternated quarter-wavelength layers as described by the following formula
\begin{center}
Incident medium / $\underbrace{\text{HLH}\cdots}_{p\text{ layers}}$ / Substrate
\end{center}
The refractive index of the semi-infinite glass substrate is denoted by $n_s$, whereas that of the semi-infinite incident medium is denoted by $n_0$. The incident medium and the substrate are non-absorbing positive-index materials. Each layer of the stack can be composed of a positive (high-index $n_H$, low-index $n_L$) or a negative index material ($-n_H$, $-n_L$), each of which should be non-absorbing.

Moreover, in this first approach we neglect the dispersion law of refractive indices for all materials under study. Though this assumption may appear too simplistic in the case of negative index materials, recently published results \cite{Jiang_2013} showed that it was possible to efficiently control this dispersion in a wide spectral range.

\subsection{Basic relations}

To determine the reflection properties of such a stack, we use the following basic formula
\begin{equation}
r=\frac{\tilde{n}_0-Y_0}{\tilde{n}_0+Y_0}=\sqrt{R}\thinspace e^{i\rho}
\label{eq:ReflectionCoefficient}
\end{equation}
where $\tilde{n}_0$ is the effective index of the incident medium, $Y_0$ is the complex admittance of the stack, $r$ is the amplitude reflection coefficient, $R$ is the corresponding reflectance and $\rho$ is the phase change at the reflection.

The effective index of a medium is given by the general relation \cite{Amra_1993}
\begin{equation}
\tilde{n}=\left\{
\begin{aligned}
&\alpha/\omega\mu_0\mu_r\quad\text{for TE polarization}\\
&\omega\epsilon_0\epsilon_r/\alpha\quad\enskip\text{for TM polarization}
\end{aligned}
\right.
\label{eq:EffectiveIndexAlpha}
\end{equation}
where $\alpha$ is defined by
\begin{equation}
\alpha^2=k^2-\sigma^2
\label{eq:alpha2}
\end{equation}
For a plane wave passing through a multilayer stack that contains $p$ layers, $\sigma$ is an invariant quantity that is defined by the angle of incidence (AOI) $\theta_0$  in the incident medium
\begin{equation}
\sigma=k_j\sin\theta_j=k_0\sin\theta_0\quad j=1,\dots,p
\end{equation}
where $j$ is the layer number, while $\alpha$ depends on the layer and is defined, in the propagating mode, by ($ j=0,1,\dots,p, p+1$)
\begin{equation}
\alpha_j=\omega\sqrt{\epsilon_0\mu_0}\sqrt{\epsilon_{r,j}\mu_{r,j}}\cos\theta_j=k_j\cos\theta_j
\end{equation}
Consequently, we can rewrite relation (\ref{eq:EffectiveIndexAlpha}) in the form 
\begin{equation}
\tilde{n}_j=\left\{
\begin{aligned}
&\frac{1}{\eta_0\mu_{r,j}}\thinspace n_j\cos\theta_j\quad\text{for TE polarization}\\
&\frac{1}{\eta_0\mu_{r,j}}\thinspace\frac{1}{n_j\cos\theta_j}\quad\text{for TM polarization}
\label{eq:EffectiveIndexTheta}
\end{aligned}
\right.
\end{equation}
where $\eta_0=\sqrt{\frac{\mu_0}{\epsilon_0}}$ is the vacuum impedance.

Relation (\ref{eq:EffectiveIndexTheta}) is independent of the type of material (PIM or NIM) within the layer because $n_j$ and $\mu_{r,j}$ are simultaneously negative in the case of negative index materials.

The computation of the $Y_0$ factor is based on the application of a recursive formula that links the admittances at two consecutive boundaries \cite{Amra_1993}
\begin{equation}
Y_{j-1}=\frac{Y_j\cos\delta_j-i\tilde{n}_j\sin\delta_j}{\cos\delta_j-i(Y_j/\tilde{n}_j)\sin\delta_j}
\label{eq:RecursiveFormula}
\end{equation}
where $\delta_j$ is the phase delay introduced by the crossing of the layer $j$. The initialization of this recursive formula occurs in the substrate where only the outgoing plane wave is present
\begin{equation}
Y_p=\Tilde{n}_{p+1}=\tilde{n}_s
\label{eq:Initialization}
\end{equation}
The phase delay $\delta_j$ is given by
\begin{equation}
\delta_j=\frac{2\pi}{\lambda}\gamma_j t_j\sqrt{\epsilon_{r,j}\mu_{r,j}}\cos\theta_j
\end{equation}
where $t_j$ is the effectivel thickness of the layer $j$ and $\gamma_j$ a binary coefficient equal to +1 (-1) for a layer that consists of a positive index (negative index) material. This last relation justifies the statement in Section \ref{sec:Introduction} and indicating that we can replace each NIM with an equivalent PIM that is characterized by a virtual negative thickness \cite{Pendry_2003}.

\subsection{Reflectance and phase spectral dependence}

All of the layers in the Bragg mirror stack are quarter-wavelength; hence, at zero AOI, we can write
\begin{equation}
t_j\sqrt{\epsilon_{r,j}\mu_{r,j}}=\frac{\lambda_0}{4}\quad\Rightarrow\quad\delta_j=\frac{\pi}{2}\gamma_j\frac{\lambda_0}{\lambda}
\end{equation}
where $\lambda_0$ is the central wavelength of the mirror. If we use a linear approximation of this last relation near $\lambda_0$, we have
\begin{equation}
\delta_j\approx\gamma_j\frac{\pi}{2}-\gamma_j\frac{\pi}{2}\thinspace\frac{x}{\lambda_0}\quad\text{with}\quad x=\lambda-\lambda_0
\end{equation}
Consequently, the recursive formula (\ref{eq:RecursiveFormula}) becomes, at the same level of approximation
\begin{equation}
Y_{j-1}\approx\frac{\tilde{n}_j^2}{Y_j}\left[1+i\frac{\pi}{2}\thinspace\frac{x}{\lambda_0}\gamma_j\left(\frac{Y_j}{\tilde{n}_j}-\frac{\tilde{n}_j}{Y_j}\right)\right]
\end{equation}
By initializing this new recursive formula using relation (\ref{eq:Initialization}), we find
\begin{equation}
Y_0=Y_0(\lambda_0)\left\{1+ix(-1)^{p-1}\frac{\pi A}{2\lambda_0}\right\}
\label{eq:LinearAdmittance}
\end{equation}
with
\begin{equation}
Y_0(\lambda_0)=\left\{
\begin{aligned}
\tilde{n}_s&\left[\frac{\tilde{n}_L^2}{\tilde{n}_H^2}\right]^q\enskip\text{for }p=2q\\
\frac{\tilde{n}_H^2}{\tilde{n}_s}&\left[\frac{\tilde{n}_H^{2}}{\tilde{n}_L^{2}}\right]^q\enskip\text{for }p=2q+1
\end{aligned}
\right.
\label{eq:Y0Lambda0}
\end{equation}
and
\begin{equation}
A=\frac{\tilde{n}_s}{\tilde{n}_H}\sum\limits_{l=0}^{p-1}\gamma_{p-l}\left[\frac{\tilde{n}_L}{\tilde{n}_H}\right]^l-\frac{\tilde{n}_H}{\tilde{n}_s}\sum\limits_{l=0}^{p-1}\gamma_{p-l}\left[\frac{\tilde{n}_H}{\tilde{n}_L}\right]^l
\label{eq:ALambda0}
\end{equation}
By combining (\ref{eq:ReflectionCoefficient}) and (\ref{eq:LinearAdmittance}), we finally obtain analytical expressions for the reflectance $R$ and the spectral derivative of the phase change at the reflection $\frac{\partial\rho}{\partial\lambda}$, both at the design wavelength $\lambda_0$
\begin{equation}
R(\lambda_0)=\left[\frac{\tilde{n}_0-Y_0(\lambda_0)}{\tilde{n}_0+Y_0(\lambda_0)}\right]^2
\end{equation}
\begin{equation}
\left.\frac{\partial\rho}{\partial\lambda}\right|_{\lambda_0}=(-1)^p\frac{\pi}{\lambda_0}\thinspace\frac{\tilde{n}_0Y_0(\lambda_0)}{\tilde{n}_0^2-Y_0^2(\lambda_0)}\thinspace A
\end{equation}

\section{Single NIM/PIM multilayer Fabry-Perot cavity}

A multilayer Fabry-Perot (FP) cavity is composed of a thin spacer (with refractive index $n_{sp}$ and thickness $t_{sp}$) surrounded by two quarter-wavelength Bragg mirrors deposited at the surface of a semi-infinite substrate. The overall round trip phase $\Phi$ of this planar cavity is defined by 
\begin{equation}
\Phi(\lambda)=2\pi\frac{\Delta}{\lambda}+\rho_{\text{US}}^-(\lambda)+\rho_{\text{LS}}^+(\lambda)
\end{equation}
where $\Delta$ is the optical path difference corresponding to the round trip of the light in the spacer layer ($\Delta=2n_{sp}t_{sp}$) and $\rho_{\text{US}}^-$ ($\rho_{\text{LS}}^+$) the phase change at the reflection on the upper mirror (lower mirror).

A Fabry-Perot resonance is defined by a central wavelength $\lambda_0$ for which the overall round-trip phase $\Phi(\lambda_0)$ is a multiple of $2\pi$ and by a spectral bandwidth $\Delta\lambda$ which is given, at leading order approximation, by
\begin{equation}
\Delta\lambda\approx\frac{2}{|\left[\frac{\partial\Phi}{\partial\lambda}\right]_{\lambda_0}|}\cdot\frac{1-\sqrt{R_{\text{LS}}^+R_{\text{US}}^-}}{(R_{\text{LS}}^+R_{\text{US}}^-)^{\frac{1}{4}}}
\end{equation}
In other words, the spectral bandwidth of such a FP resonance can become extremely large if the linear dependence of the overall round-trip phase in the cavity
\begin{equation}
\left[\frac{\partial\Phi}{\partial\lambda}\right]_{\lambda_0}=-\frac{4\pi n_{\text{sp}}e_{\text{sp}}}{\lambda_0^2}+\left.\frac{\partial\rho_{\text{US}}^-}{\partial\lambda}\right|_{\lambda_0}+\left.\frac{\partial\rho_{\text{LS}}^+}{\partial\lambda}\right|_{\lambda_0}
\label{eq:PhiLinearDependence}
\end{equation}
is equal or very close to zero.

To fulfill this condition and thus obtain a resonant behavior over a very large spectral range, it is absolutely required that at least \textbf{one of the cavity mirrors includes NIM layers}, which is the only way to obtain a \textbf{positive} linear dependence of the phase change at reflection \cite{Lequime_2013_2}.

To determine the optimal design of such a \textbf{\textit{white} multilayer Fabry-Perot cavity}, we systematically investigate the variation of $\left[\frac{\partial\Phi}{\partial\lambda}\right]_{\lambda_0}$ of all the symmetric multilayer stacks described by the general formula
\begin{center}
Air / H$\text{(LH)}^q$ 2L $\text{(HL)}^q$H / Glass
\end{center}
when the refractive index $n_H$ of the high-index material varies in the range between 2.00 and 3.00, while the other refractive indices remain constant ($n_0$ = 1.00, $n_s$ = 1.52 and $n_L$ = 1.48). In the previous stack formula, $q$ is an integer equal to 1, 2 or 3, while H and L represent either PIM (H, L) or NIM ($\bar{H}$, $\bar{L}$) quarter-wavelength layers. The calculus is performed for zero AOI.

For each stack formula that exhibits a cancellation of the spectral dependence of the overall round-trip phase at the design wavelength $\lambda_0$, we thus calculate its spectral transmittance $T(\lambda)$, and the variation of the ratio between the square modulus $|\vec{\mathcal{E}}(Z)|^2$ of the electric field within the stack and the square modulus of the incident field $|\vec{\mathcal{E}}_0^+|^2$, where $Z$ is the coordinate along a vertical axis perpendicular to the substrate plane, whose origin is taken to be at the top of the stack.

These two quantities are computed using a recursive relation between the electric fields at the boundaries, completed by the related initialization condition
\begin{equation}
\left\{
\begin{aligned}
&\vec{\mathcal{E}}_{j-1}=\left[\cos\delta_j-i(Y_j/\tilde{n}_j)\sin\delta_j\right]\thinspace\vec{\mathcal{E}}_{j}\\
&\vec{\mathcal{E}}_{0}=(1+r)\vec{\mathcal{E}}_{0}^+
\end{aligned}
\right.
\end{equation}
and extended in the thickness of each layer by
\begin{multline}
\vec{\mathcal{E}}_{j}(z)=\left[\cos\delta_j(z)-i(Y_j/\tilde{n}_j)\sin\delta_j(z)\right]\thinspace\vec{\mathcal{E}}_{j}\\
\text{with}\enskip\delta_j(z)=\alpha_j(e_j-z)
\end{multline}

\begin{figure}[H]
\centerline{\includegraphics[width=0.8\columnwidth]{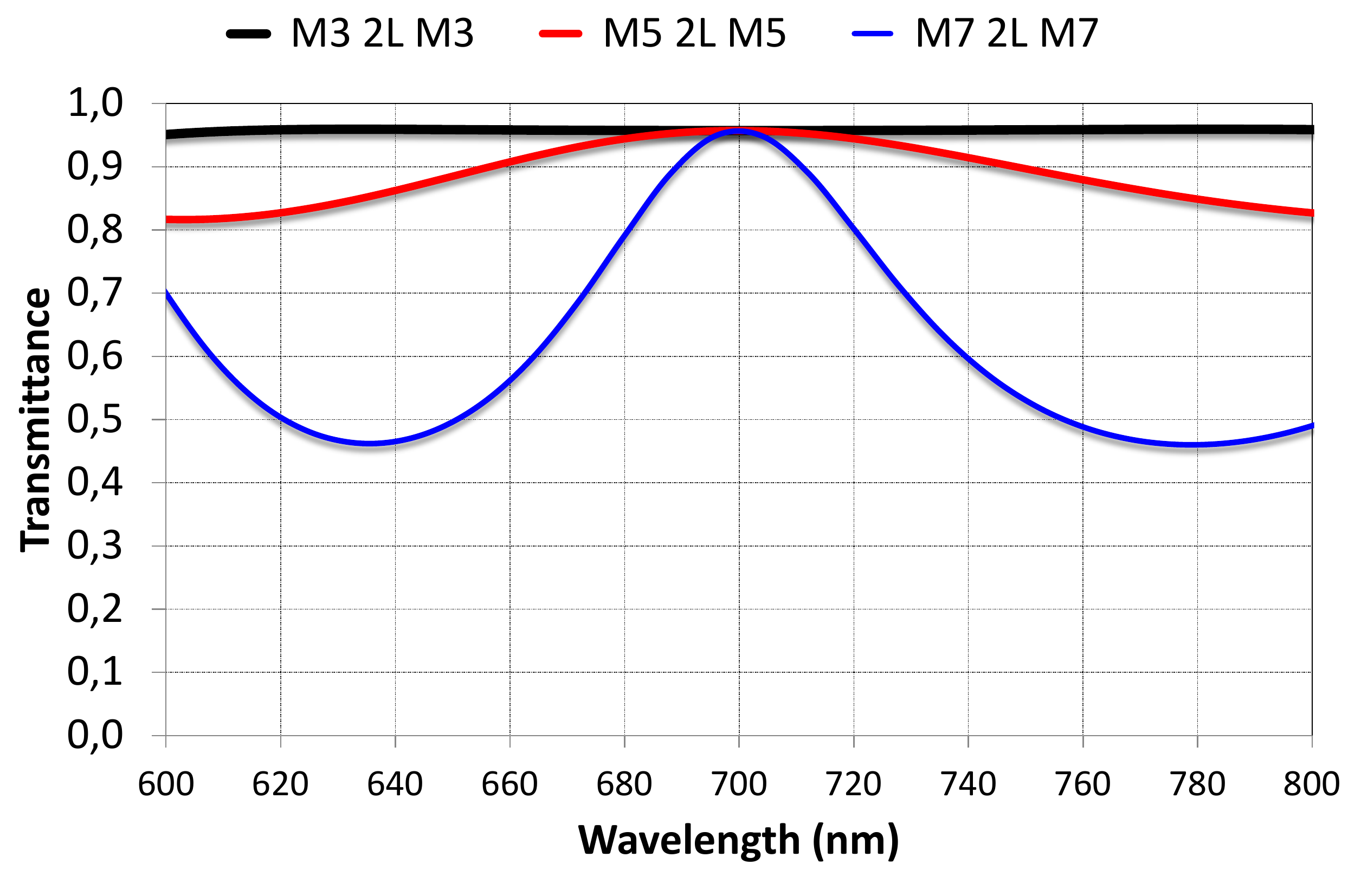}}
\caption{Spectral transmittance of three optimized \textit{white} multilayer Fabry-Perot configurations. Black line: $\bar{H}\bar{L}\bar{H}$ 2L $\bar{H}\bar{L}\bar{H}$, $n_H$ = 2.61; red line: $\bar{H}\bar{L}\bar{H}$L$\bar{H}$ 2L $\bar{H}$L$\bar{H}\bar{L}\bar{H}$, $n_H$ = 2.19; blue line: $\bar{H}\bar{L}\bar{H}\bar{L}\bar{H}$L$\bar{H}$ 2L $\bar{H}$L$\bar{H}\bar{L}\bar{H}\bar{L}\bar{H}$, $n_H$ = 2.37.}
\label{fig:WhiteFP}
\end{figure}

To quantify and compare the strength and the spectral width of the resonance behavior of these various Fabry-Perot configurations, we introduce the merit factor $M$
\begin{equation}
M=\frac{1}{\lambda_2-\lambda_1}\int_{\lambda_1}^{\lambda_2}\frac{|\vec{\mathcal{E}}_{\text{sp}}(e_{\text{sp}}/2,\lambda)|^2}{|\mathcal{E}_0^+|^2}\thinspace d\lambda
\end{equation}
where $\lambda_1$ and $\lambda_2$ are the limits of the selected spectral range (here, $\lambda_1$ = 600 nm and $\lambda_2$ = 800 nm, cf. Fig. \ref{fig:WhiteFP}), and $\vec{\mathcal{E}}(e_{\text{sp}}/2,\lambda)$ is the spectral dependence of the amplitude of the electric field in the middle of the spacer.

The Table \ref{tab:FabryPerotConfigurations} summarizes the results provided by this systematic screening and gives, for each optimized configuration, the stack formula, the refractive index of the H-layer, the value of the merit factor $M$ and the ratio $Q$ between this merit factor and that of the corresponding all-PIM cavity (for all these cavities, the design wavelength $\lambda_0$ is equal to 700 nm).
\begin{table}[htbp]
\caption{\label{tab:FabryPerotConfigurations}Main features of three optimized symmetric \textit{white} Fabry-Perot configurations}
\begin{ruledtabular}
\begin{tabular}{lccc}
FP stack formula & $n_H$ & $M$ & $Q$\\
$\bar{H}\bar{L}\bar{H}$ 2L $\bar{H}\bar{L}\bar{H}$ & 2.61 & 5.70 & 3.4\\
$\bar{H}\bar{L}\bar{H}$L$\bar{H}$ 2L $\bar{H}$L$\bar{H}\bar{L}\bar{H}$ & 2.19 & 5.83 & 4.3\\
$\bar{H}\bar{L}\bar{H}\bar{L}\bar{H}$L$\bar{H}$ 2L $\bar{H}$L$\bar{H}\bar{L}\bar{H}\bar{L}\bar{H}$ & 2.37 & 15.3 & 10.5\\
\end{tabular}
\end{ruledtabular}
\end{table}
Figure \ref{fig:WhiteFP} shows the spectral transmittance of these three optimized Fabry-Perot configurations; the most attractive wide-band behavior seems provided by the first two listed (M3 2L M3 and M5 2L M5).

\begin{figure}[htbp]
\centerline{\includegraphics[width=0.8\columnwidth]{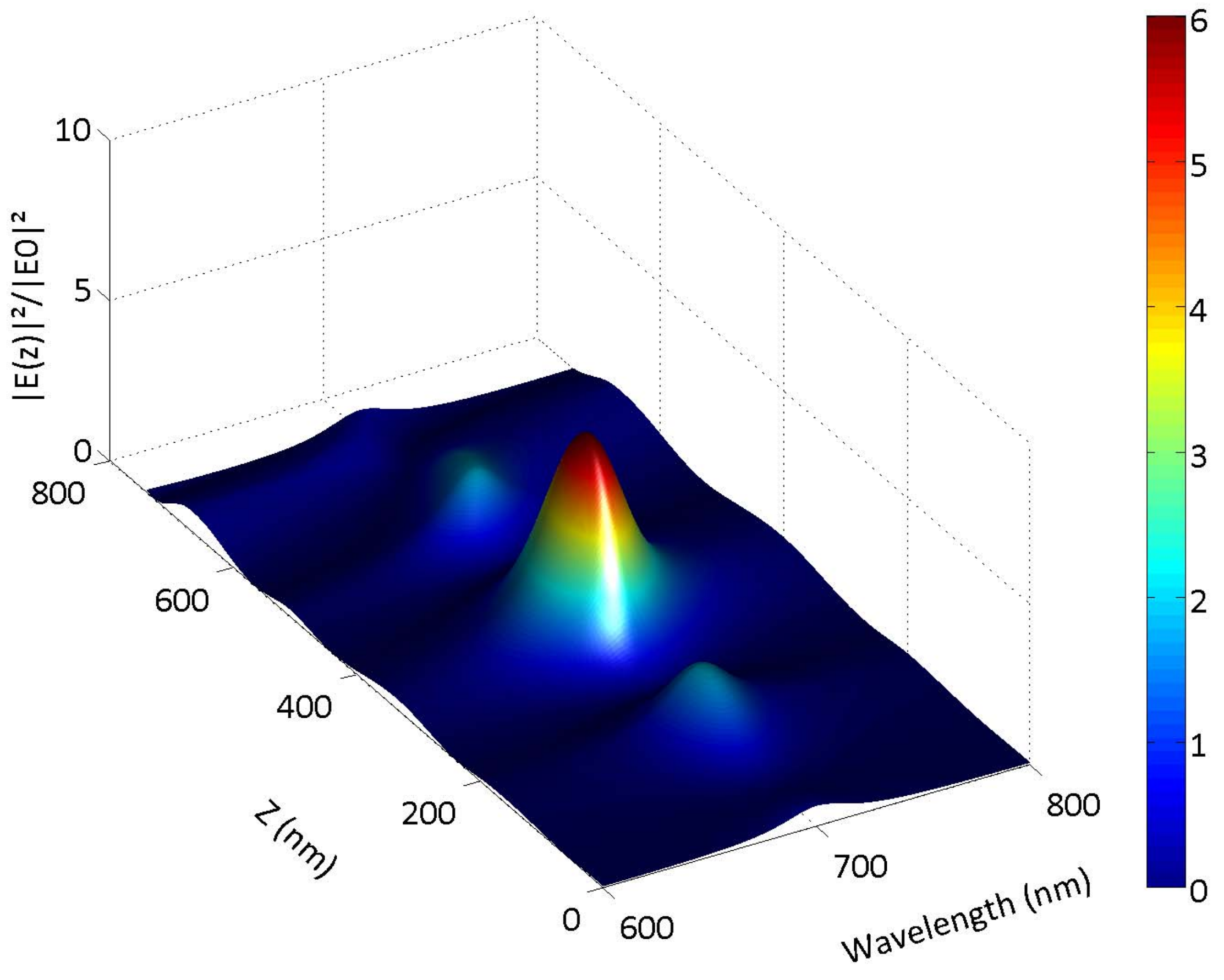}}
\caption{Spectral dependence of the normalized square modulus of the electric field within the thickness of an HLH 2L HLH standard Fabry-Perot ($n_H$ = 2.61, $n_L$ = 1.48).}
\label{fig:PIMM32LM3}
\end{figure}

\begin{figure}[htbp]
\centerline{\includegraphics[width=0.8\columnwidth]{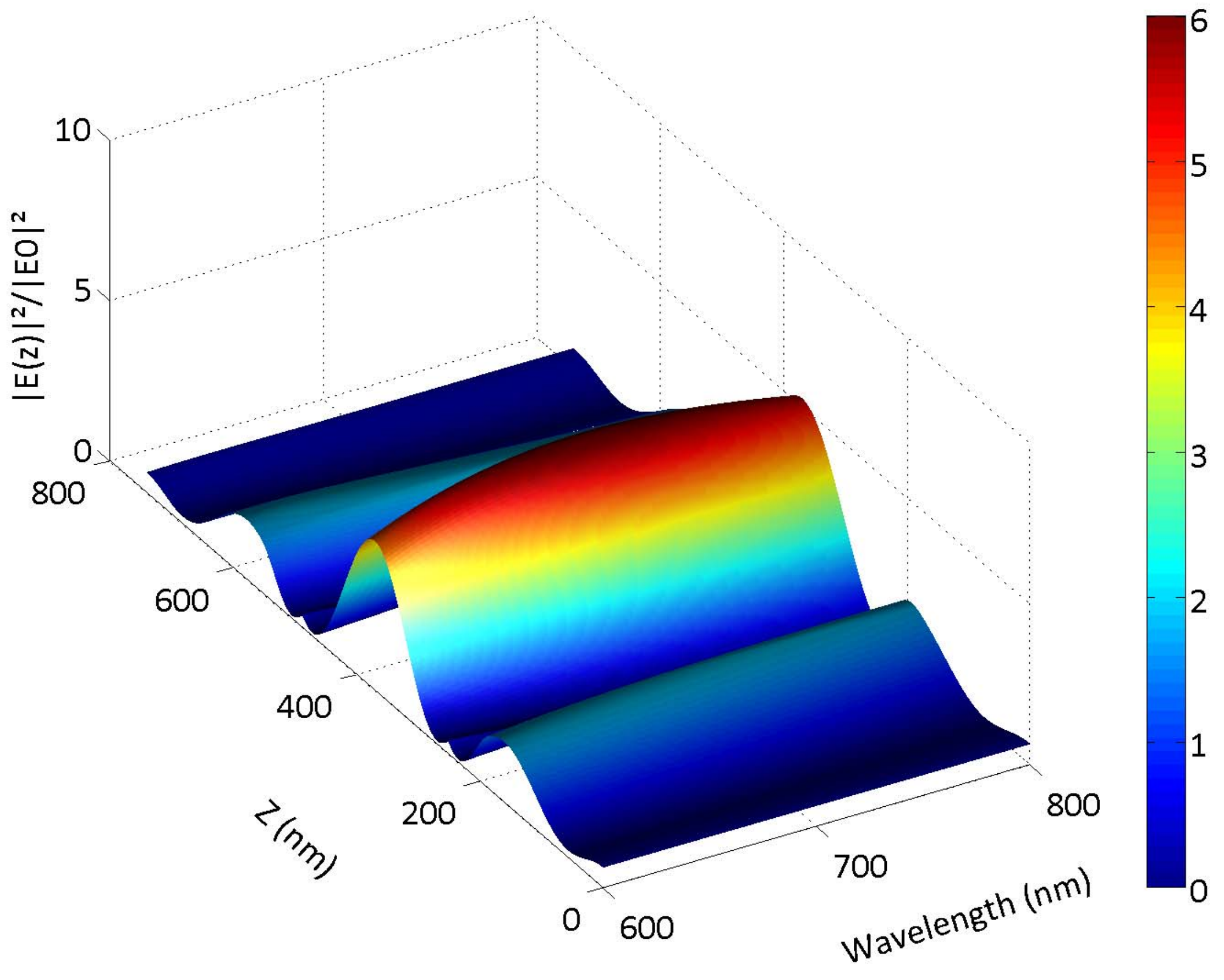}}
\caption{Spectral dependence of the normalized square modulus of the electric field within the thickness of an $\bar{H}\bar{L}\bar{H}$ 2L $\bar{H}\bar{L}\bar{H}$ Fabry-Perot configuration (the refractive indices of the NIM layers are opposite to those of the standard configuration).}
\label{fig:NIMPIMM32LM3}
\end{figure}

The variations in the normalized square modulus of the electric field within these stacks are presented in Fig. \ref{fig:PIMM32LM3} for a standard all-PIM M3 2L M3 configuration ($n_H$ = 2.61, $n_L$ = 1.480) and in Fig. \ref{fig:NIMPIMM32LM3} for the optimized symmetric NIM/PIM Fabry-Perot configuration ($\bar{H}\bar{L}\bar{H}$ 2L $\bar{H}\bar{L}\bar{H}$). The use of NIM layers provides, as expected, a spectacular increase of the spectral range in which resonant behavior is achieved.


\section{Conclusion}
We have shown how negative index materials can be used to design a novel type of planar multilayer cavity, which we propose to call a \textit{white} Fabry-Perot cavity, and which is resonant over a very large range of wavelengths in comparison to standard devices. More detailed analysis demonstrates that the choice of the refractive index $n_H$ of each optimized configuration is quite tolerant, which is advantageous for possible practical implementations.

This type of planar multilayer cavity can be very effective in optimizing the light-extraction efficiency of surface emitters \cite{Lequime_2010}, such as small-molecule Organic Light-Emitting Diodes (OLEDs).

Moreover, by slightly modifying the features of the optimized M5 2L M5 Fabry-Perot cavity described in the previous section (centering wavelength 600 nm, modified formula $\bar{H}$L$\bar{H}$L$\bar{H}$ 2L $\bar{H}$L$\bar{H}$L$\bar{H}$, $n_H$ = -2.24 for all high-index NIM layers) and by stacking 3 identical cavities linked by a PIM low-index quarter-wavelength layer, we are able to obtain a filter with a nice rectangular profile, a bandwidth of approximately 90 nm, and a high level of rejection throughout the entire remaining spectral range (approximately -30 dB). The addition of negative index materials to the data-base of standard thin-film software allows us to define optimized designs for many filtering applications.

Further efforts will be devoted to an in-depth analysis of all these exciting potential topics.

\end{document}